\documentclass{emulateapj}

\usepackage{amsmath}	
\usepackage{amssymb}	

\newcommand{\tb}{\theta_B}
\newcommand{\Ng}{N_{\mathrm{coinc}}}
\newcommand{\Nb}{N_{\mathrm{GW}}}
\newcommand{\g}{_\mathrm{GRB}}

\newcommand{\Ab}{\mathcal{A}_\mathrm{GW}}
\newcommand{\Ag}{\mathcal{A}_\mathrm{coinc}}
\newcommand{\diff}{\text{d}}
\newcommand{\Lmax}{L_{\mathrm{max}}}
\newcommand{\focc}{f_{\mathrm{vis.}}}

\newcommand{\epscounthundred}{12\%}
\newcommand{\sigmacounthundred}{3}
 
\newcommand{\epspehundred}{9.6\%} 
\newcommand{\sigmapehundred}{2.7} 
\newcommand{\epscountfive}{51\%}

\newcommand{\epspefive}{43\%}

\received{XXX}
\revised{XXX}
\accepted{XXX}

\shorttitle{Counting on Short Gamma-Ray Bursts}

\begin{document}

\title{
Counting on Short Gamma-Ray Bursts: Gravitational-Wave Constraints of Jet Geometry
}


\author{Amanda Farah}
\affiliation{Department of Physics,
University of Chicago,
Chicago, IL 60637, USA}
\affiliation{Kavli Institute for Cosmological Physics,
University of Chicago,
Chicago, IL 60637, USA}

\author{Reed Essick}
\affiliation{Kavli Institute for Cosmological Physics,
University of Chicago,
Chicago, IL 60637, USA}

\author{Zoheyr Doctor}
\affiliation{Department of Physics,
University of Oregon,
Eugene, OR 97403, USA}
\affiliation{Department of Physics,
University of Chicago,
Chicago, IL 60637, USA}
\affiliation{Kavli Institute for Cosmological Physics,
University of Chicago,
Chicago, IL 60637, USA}
\affiliation{Enrico Fermi Institute,
University of Chicago,
Chicago, IL 60637, USA}

\author{Maya Fishbach}
\affiliation{Department of Astronomy and Astrophysics,
University of Chicago,
Chicago, IL 60637, USA}
\affiliation{Kavli Institute for Cosmological Physics,
University of Chicago,
Chicago, IL 60637, USA}

\author{Daniel E. Holz}
\affiliation{Department of Physics,
University of Chicago,
Chicago, IL 60637, USA}
\affiliation{Department of Astronomy and Astrophysics,
University of Chicago,
Chicago, IL 60637, USA}
\affiliation{Kavli Institute for Cosmological Physics,
University of Chicago,
Chicago, IL 60637, USA}

\begin{abstract}
The detection of GW170817 in gravitational waves and gamma rays revealed that at least some short gamma-ray bursts are associated with the merger of neutron-star compact binaries.
The gamma rays are thought to result from the formation of collimated jets, but the details of this process continue to elude us.
One fundamental observable is the emission profile of the jet as a function of viewing angle.
We present two related methods to measure the effective angular width, $\tb$, of short gamma-ray burst (sGRB) jets using gravitational wave and gamma-ray data, assuming all sGRBs have the same angular dependence for their luminosities.
The first is a counting experiment, where we combine the known detection thresholds of the LIGO/Virgo and Fermi Gamma Ray Burst Monitor detectors to infer parameters of systems that are detected in gravitational waves. 
This method requires minimal knowledge about each event, beyond whether or not they were detected in gamma-rays.
The second method uses additional information gleaned from the  gravitational-wave and electromagnetic data to estimate parameters of the source, and thereby improve constraints on jet properties.
We additionally outline a model-independent method to infer the full jet structure of sGRBs using a non-parametric approach.
Applying our methods to GW170817, we find only weak constraints on the sGRB luminosity profile, with statistical uncertainty dominating differences between models.
We also analyze simulated events from future observing runs, and find that
with 5 and 100 BNS detections, the counting method constrains the relative uncertainty in $\tb$ to within $\epscountfive$ and $\epscounthundred$, respectively.
Incorporating gravitational-wave parameter estimation would further tighten these constraints to $\epspefive$ and $\epspehundred$.
In the limit of many detections, incorporating parameter estimation achieves only marginal improvements; we conclude that the majority of the information about jet structure comes from the relative sensitivities of gravitational-wave and gamma-ray detectors as encoded in simple counting experiments.
 \end{abstract}



\section{Introduction} 
\label{sec:intro}
Because of their scarcity, irregularity, and brevity, very little is known about the origin and formation mechanisms of short gamma-ray bursts (sGRBs).
Though it was postulated for decades that sGRBs could result from the mergers of binary neutron star (BNS) systems \citep[e.g.][]{Eichler_1989,Narayan_1992,Fong_2010,Church_2011} or neutron star-black hole systems (see \cite{2014ARA&A..52...43B} for a recent review), the  joint gamma-ray and gravitational wave detection of GW170817 confirmed this association observationally for the first time \citep{170817_grb_ligo, 0817_ligo, Wu2019}.
Indeed, this breakthrough was only possible by simultaneously observing the multimessenger sky with the \textit{Fermi} Gamma-Ray Burst Monitor (GBM,~\cite{2009ApJ...702..791M})~and the advanced LIGO \citep{aligo}~and Virgo~\citep{2015CQGra..32b4001A} gravitational-wave (GW) detectors.
However, there still remain many open questions about the mechanism by which gamma rays are generated. 
One of the simplest and most important of these is the angular structure of the gamma-ray emission, a fundamental property of sGRBs which is sensitive to the underlying astrophysics of the merger \citep{2014ApJ...784L..28N,2005A&A...436..273A,2015PhR...561....1K}.

Although the specific angular geometry of sGRB jets is unknown, there are several features believed to be common to these phenomena.
First, sGRB jets are thought to be launched from the poles of the remnant left over after the coalescence of two neutron stars or a neutron star and a black hole.
The precise mechanism by which the jet is launched is still unknown (\cite{2014ARA&A..52...43B} but see also \cite{2015ApJS..218...12L} and references therein), but it is generally believed that there are symmetric polar outflows of highly relativistic material that travel parallel to the binary system's orbital angular momentum.
Furthermore, jets are thought to be collimated and roughly axisymmetric, emitting preferentially in a narrow opening angle due to a combination of outflow geometry and relativistic beaming.
Importantly, the angular dependence of the jet luminosity is very uncertain. We assume it decreases monotonically at large viewing angles ({off-axis}) compared to lines of sight nearly aligned with the progenitor system's angular momentum ({on-axis}). 
This means that the majority of sGRBs are only detectable if they are aligned within a narrow window around our line of sight, although off-axis detection is still plausible if the source is at sufficiently low redshift \citep{2017MNRAS.471.1652L,2012ApJ...746...48M}.
This could explain why GRB 170817A was highly sub-luminous, although it has been argued that it might instead be a member of a separate sub-luminous population of sGRBs~\citep{2016arXiv160603043S,170817_grb_ligo}.
GW radiation from the sGRB progenitor, although still preferentially emitted along the orbit's angular momentum, has a much shallower angular dependence.
These systems are therefore detectable in GWs at much larger viewing angles than sGRBs.
We investigate the interplay between the different angular scales in sGRB and GW emission profiles, following up on previous counting experiments (e.g. \cite{2014ApJ...784L..28N,2013PhRvL.111r1101C}) and investigating the impact of additional information beyond the relative sensitivities of GW and gamma-ray detectors.

We present two methods to infer the geometrical properties of sGRB emission by relating parameters that can be extracted from the GW signal, such as the inclination and redshift of the system, to the number of sGRBs detected.
We also make a prediction for the constraining power of these methods, and show that, in an optimistic scenario of 100 BNS detections and a tophat jet structure, the beaming angle will be constrained to within $\sigmapehundred^\circ$ if $\leq 20$ of these events have associated sGRBs.
In addition, we find that the ability to  jet structure are relatively model independent.
Incorporating GW measurements of the systems' inclinations will initially help constrain jet structure, although the basic counting experiment will produce nearly equivalent constraints by the time modeling systematics dominate over statistical uncertainty.

The framework presented here is complementary to current methods which consider only EM data, and  infer beaming angles for each sGRB through radio afterglow observations \citep{Fong2015,2016A&A...594A..84G}.
Radio afterglow measurements employ the fact that a jet break occurs when the relativistic Lorentz factor ($\Gamma$) approaches $\sim \tb^{-1}$, resulting in a characteristic steepening of the light curve \citep{2018ApJ...859..160W,2018ApJ...857..128J}.
Given a model for the evolution of $\Gamma$ over time, \citep[e.g.][]{1976PhFl...19.1130B,2014ApJ...796...30S} one can then deduce the beaming angle using the time between gamma-ray emission and jet break.
Thus, an independent measurement of the beaming angle would enable progress on the reverse problem, allowing for insights on the energetics of the burst \citep{Frail2001,2014ApJ...784L..28N,2017JHEAp..13....1Y}.
Instead of inferring jet structure parameters for individual sGRBs \citep{2018MNRAS.478.4128G,2018NatAs...2..751L,2018ApJ...858...65L,2018ApJ...852L...1Z,2016ApJ...833...88L,2016ApJ...822L..14Z}, we consider the population of sGRBs that are accompanied by GWs and constrain the beaming angle with observations of multiple sGRBs. 

Other studies have investigated similar relationships.
For example, \cite{kentaro} take a sample of sGRBs with known luminosities and infer a rate of BNS mergers for several choices of luminosity functions. 
Assuming a universal structured jet in the form of a broken power law, they constrain the jet parameters based on GW170817 \citep{0817_ligo} and their sGRB sample.
Using these two measurements, they infer an expected rate of \emph{coincident} gamma-ray bursts and GW detections. 

We solve a related problem.
Through a simple counting experiment that expands on the work in ~\cite{2013PhRvL.111r1101C}, we constrain sGRB jet parameters by using the number of sGRBs detected in coincidence with BNS mergers detected in GWs.
Instead of predicting the rate of coincident detections based on a model of the jet structure, we study how well jet structures can be constrained given a set of coincident detections. 
To do this, we sample BNS systems from a physically motivated redshift, inclination, and maximum luminosity distribution.
Then, taking advantage of the redshift and inclination dependence of both GW detector and GBM sensitivities, we determine the likelihood of detecting BNSs in GWs with and without associated sGRBs as a function of $\tb$.
We examine various 2-parameter jet structure models in this way.
Additionally, we show how to easily extend this analysis to include posterior distributions from GW data when available.

\cite{sylvia} and \cite{fergus} have recently presented similar studies. We highlight a number of important difference with their work.
First, neither \cite{sylvia} nor \cite{fergus} consider selection effects, whereas this work shows that knowledge of selection effects lead to constraints on the jet width that are comparable to those obtained from GW parameter estimation. 
Secondly, both \cite{sylvia} and \cite{fergus} parameterize the fluence of gamma-rays in terms of the jet structure models that they consider, whereas this work only considers luminosity to the extent that events are above the GBM threshold. The methods in \cite{sylvia} and \cite{fergus} therefore require additional information from the gamma-ray detection, beyond what is assumed in the methods we present.

Additionally, \cite{fergus} focuses mainly on model selection between two jet structures.
It discusses the number of events it would take to distinguish between jet structures.
We make no such attempt, but show that different jet structures can produce different constraints. 
\cite{sylvia}
assumes a joint detection or the availability of a fluence upper limit, and employs parameter estimation to obtain posteriors for gravitational wave and electromagnetic counterpart parameters. This necessitates simulating 200 BNS merger events, specifying waveform models, and injecting these events into LIGO detector networks. 

The work presented here takes a simpler approach, where the posterior on jet width given the number of GW and sGRB detections is analytically derived and then calculated by sampling from priors on inclination and redshift.
To incorporate parameter estimation information, one directly applies posterior samples from the publicly available gravitational wave transient catalogs \citep{gwtc}.
Perhaps surprisingly, this straightforward approach appears to be as effective as those presented in \cite{sylvia} and \cite{fergus}, as all three studies determine the beaming angle with errors of less than $\sim 10^\circ$ after 100 events.

We make several simplifying assumptions for the physical properties of sGRBs as well as their detection criteria.
First, we only consider detections from GBM, which is sensitive to $\sim$ 70\% of the sky (the other $\sim30\%$ is occulted by the Earth) and only observes for 85\% of the time.
The parts of the sky to which GBM and LIGO/Virgo are sensitive at any moment are not believed to be correlated as the typical lock duration of GW interferometers ~\citep{detectorstatus} is much longer than Fermi's orbital period  \citep[$\sim96$ min;][]{2009ApJ...702..791M}.
Sky coverage from other instruments like \textit{Swift}-BAT~\citep{2004ApJ...611.1005G} and Konus-\textit{Wind}~\citep{1995SSRv...71..265A} are negligible in comparison.

Second, we assume that matter ejected from a BNS merger emits radiation in the form of two axisymmetric jets of gamma-rays which are aligned with the system's angular momentum.
Thus, the inclination $\iota$ of the BNS relative to our line of sight is a perfect proxy for viewing angle, and we will make no distinction between the two. 

Last, and least conventional, is our assumption that sGRBs have a universal jet structure.
We choose structures that monotonically decrease with increasing viewing angle and are described by only two free parameters.
A universal jet structure $\mathcal{L}(\iota; \tb, \Lmax)$ implies that the observed isotropic equivalent luminosity of a given sGRB depends only on the viewing angle ($\iota$) of the observer, up to an overall scaling constant ($\Lmax$).
We consider three functional forms for the luminosity of the jet as function of viewing angle: a tophat, a broken power law, and a Gaussian.
In all three jet models, the two parameters are the width of the jet ($\tb$) and the luminosity at the very center of the jet ($\Lmax$).
While not guaranteed to be the case, this assumption is motivated by \cite{Wu2019} who argue that all cosmological sGRBs have the same afterglow, and differences in afterglow are due to different viewing angles and distances.
It is thus not unreasonable to postulate that the same might be true for the prompt gamma-ray emission, as it is certainly plausible that the prompt emission structure is related to the afterglow structure.
Additionally, \cite{Perna2003} show that a universal structured jet model of sGRBs is consistent with the observed distribution of viewing angles.
\cite{Nakar2004} argue that this statement is not valid for the 2D distribution of viewing angles and redshift, but note that any disagreement might be due to inhomogeneity in sGRB observations or selection effects.
Thus, it is not unreasonable to assume a universal structure for sGRB jets with the caveat that, if this assumption does not hold, any measurement of jet parameters that relies on multiple detections would be invalid.
We investigate the implications of these assumptions for future jet structure constraints.

This paper is organized as follows.
Section \ref{sec:methods} details the method used to infer jet parameters from GW-only and joint sGRB-GW detections of BNS mergers, and Section~\ref{sec:pe} explains how one would extend this method to include information from GW parameter estimation.
Section \ref{sec:results} summarizes the results and provides predicted constraints from future detections.
We conclude by discussing this method's limitations and outlining other possible approaches in Section~\ref{sec:conclusions}.


\section{Methods}
\label{sec:methods}
We carry out a Bayesian analysis to infer the width of sGRB jets from a BNS merger for three structured jet models.
There are several naming conventions in the field, so we first define our parameters.
	\begin{itemize}
    	\item $\Nb$ and $\Ng$ are the number of BNS mergers detected in GWs and the number of sGRBs detected by GBM in coincidence with a GW detection, respectively. Therefore, $0\leq\Ng\leq\Nb$.
	    \item $\mathcal{L}(\iota;\theta_B,\Lmax)$ is the \textit{intrinsic} luminosity as a function of viewing angle for \textit{all} sGRBs. This is the function that describes the jet structure. We assume there is a universal angular dependence for all sGRBs, so that $\mathcal{L}/\Lmax$ is the same for every jet.
	    \item $\tb$ is the width of the jet. For a tophat (Equation~\ref{eq:tophat}), this is the half opening angle, beyond which no light is emitted. For $\tb=\pi/2$, each jet covers half the sky, and the gamma-rays are emitted isotropically. For a Gaussian jet (Equation~\ref{eq:gauss}), $\tb$ is the standard deviation of the angular profile, and for a broken power law (Equation~\ref{eq:broken_plaw}), $\tb$ is the inclination at which the jet structure function is truncated, i.e. where the luminosity is set to zero. Note that the meaning of $\tb$ can be quite different for each jet structure model and one must take care when comparing model-dependent statements.
	    \item $\Lmax$ is the luminosity at the center of the jet and can be thought of as the overall normalization of the jet structure. In this work, we draw $\Lmax$ from a Gaussian distribution with mean $2 \times 10^{52}$ erg/s and standard deviation $2\times 10^{51}$ erg/s. This distribution was chosen to be consistent with the majority of observed sGRBs with known redshift~(see Figure 4 of \cite{170817_grb_ligo}). Note that even though the angular dependence of the jet luminosity is assumed to be universal, drawing $\Lmax$ from a distribution allows for variations in the overall luminosity of different sGRBs, which could be due to different energy reservoirs in the remnants or different component masses in the progenitor systems.
	    \item $L_{\mathrm{iso}}$ is the isotropic equivalent luminosity of a sGRB. Isotropic energetics are calculated under the assumption that the source emits isotropically, so that the luminosity in the observer's direction is the luminosity everywhere. This is often considered an upper bound on the true total energetics if one assumes that the GRB is observed at the brightest part of the jet (\S 6.1 of \cite{170817_grb_ligo}), although this is only true for a tophat jet structure.
	    \item 
	    $\iota$ is the viewing angle, or the inclination of the system relative to our line of sight. It is defined as the angle between our line of sight to the binary system and the system's angular momentum.
	    \item $q$ is the coincident fraction, defined as the fraction of BNS mergers detected in GWs that also have an associated short gamma-ray burst detected by GBM. Given the sensitivities of both detectors, the electromagnetic parameters of sGRBs, as well as an assumed distribution of binary neutron star mergers in redshift and inclination, we calculate the coincident fraction in the limit of infinite detections:
\begin{equation}
    q \equiv \lim_{\Nb\to\infty}\frac{\Ng}{\Nb} \nonumber
\end{equation}
        We then use this to infer a posterior distribution of the jet opening angle $\tb$ for a given jet structure $\mathcal{L}(\iota;\theta_B,\Lmax)$ based on a finite number of detections.
	\end{itemize}


\subsection{Bayesian Formulation} 
\label{sec:binom}

Our goal is to constrain the effective angular width of sGRB jets, $\tb$.
We calculate the posterior probability $p(\tb|\Ng, \Nb)$ based on the number of sGRB and GW detections. 

\begin{equation}
    p(\tb|\Ng, \Nb) = \frac{p(\Ng,\Nb|\tb)p(\tb)}{p(\Ng,\Nb)}
\end{equation}
We assume a uniform prior in $\tb$, so the posterior is proportional to the likelihood.
Note that the coincident fraction, $q$ can be computed from $\tb$, so we can rewrite the likelihood as $p(\Ng,\Nb|q(\tb))$.
Because the detection of a GRB is a Boolean outcome, this likelihood is a binomial distribution with success fraction $q$, number of trials $\Nb$, and number of successes $\Ng$.
Replacing the likelihood and priors with their functional forms, we obtain

\begin{multline}
    p(\tb|\Ng, \Nb) = \\
    \frac{q(\tb)^{\Nb}(1-q(\tb))^{\Nb-\Ng} }{\int\diff \tb\, q(\tb)^{\Nb}(1-q(\tb))^{\Nb-\Ng}}
    \label{eq:posterior}
\end{multline}
This shows that, for example, lower $\Ng$ and higher $\Nb$ correspond to posterior support in a narrow region around small $q$ and therefore small $\tb$.
This, then, reduces the problem of constraining the jet width to one of calculating the coincident fraction as a function of $\tb$ for a given jet structure.


\subsection{Calculating the Coincident Fraction}
\label{sec:q}

\begin{figure}
    \centering
    \includegraphics[width=\columnwidth]{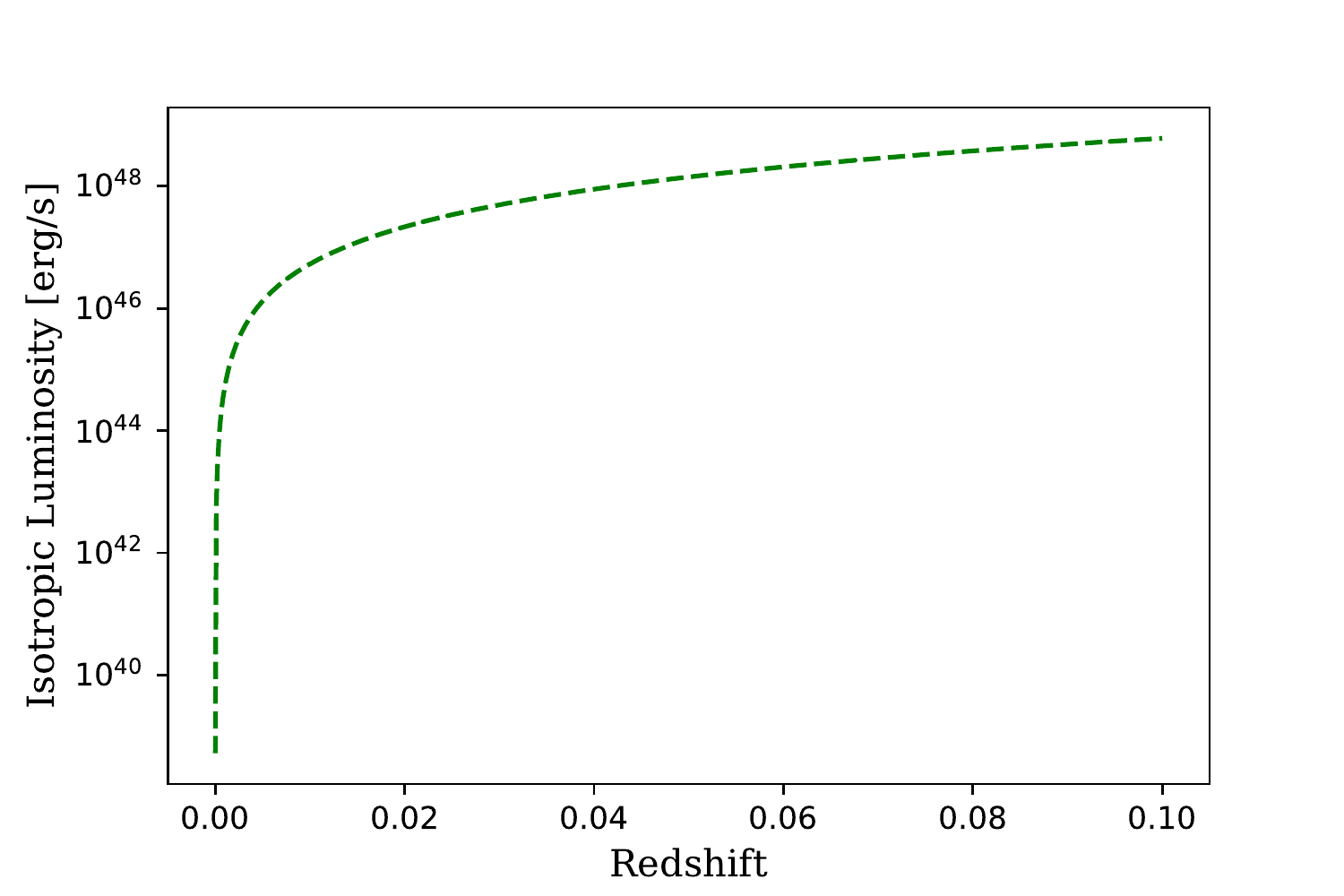} 
    \caption{
        Approximate GBM isotropic luminosity detection threshold as a function of redshift, $L_{iso,thr}$ reproduced from Figure 4 of~\cite{170817_grb_ligo}.
        Note that this threshold increases slowly for $z \gtrsim 0.01$.
    }
    \label{fig:sens_curve} 
\end{figure}

The coincident fraction ($q$) is defined as the probability of detecting an sGRB given a BNS merger detection:

\begin{equation}
    q = p(\Ag=1|\Ab=1)
\end{equation}
where $\Ag$ ($\Ab$) is a Boolean indicator that represents whether or not the system was detected in gamma-rays (GWs).
While individual detections depend on parameters such as source inclination and distance, we ultimately want $q$ only as a function of the beaming angle $\tb$ and the intrinsic maximum luminosity of the jet $\Lmax$.
Thus, to obtain $q(\tb)$, we marginalize over the unknown inclination angle $\iota$ and redshift $z$ of a given system. 

\begin{equation}
    q = \int \diff z \diff \iota\, p(\Ag=1|\iota,z)p(\iota,z|\Ab=1) \label{eq:marginalized_q}
\end{equation}
GBM detects an event when the observed flux is above a threshold, and a GRB's flux can be calculated from its equivalent isotropic luminosity $L_{\mathrm{iso}}$ and distance from Earth.
Thus, GBM's flux threshold can be converted to an isotropic luminosity threshold as a function of redshift, as shown in Figure~\ref{fig:sens_curve}.
We therefore write $p(\Ag|\iota,z)$ as a function of $L_{\mathrm{iso}}$ and $z$,

\begin{equation}
    p(\Ag=1|\iota,z)  = \focc \Theta(L_{\mathrm{iso}}(z,\iota) - L_{\mathrm{iso,thr}}(z)) 
    \label{eq:pdet_grb_heaviside}
\end{equation}
where $\Theta(x) $ is the Heaviside step function and $L_{\mathrm{iso,thr}}(z)$ is GBM's isotropic luminosity threshold at redshift $z$.
$\focc$ is the fraction of the sky to which GBM is sensitive.
This is taken to be $0.7$ since the Earth occults $\sim 30\%$ of GBM's field of view.
However, this number can be modified to account for time spent in the South Atlantic Anomaly, as well as additional time for slewing and safe holding.
For example, \cite{Burns_2016} calculates a time-averaged sky fraction of $\sim 0.6$ \citep{Burns_2016}. 
$L_{\mathrm{iso}}(z,\iota)$ can be calculated by knowing the jet's luminosity as a function of viewing angle, $\mathcal{L}(\iota;\tb,\Lmax)$.
Since isotropic energetics are calculated by assuming that the luminosity in the observer's direction is the luminosity everywhere, the observed luminosity would be the same for observers at all viewing angles.
Then,

\begin{equation}
    L_{\mathrm{iso}} = 4 \pi \mathcal{L}(\iota;\tb,\Lmax)
    \label{eq:Liso}
\end{equation} 

We assume a universal jet structure, so $\mathcal{L}$ is known and the same for all sGRBs.
We consider tophat, ($\mathcal{L}_T$), Gaussian ($\mathcal{L}_G$), and broken power law ($\mathcal{L}_P$) jets:

\begin{align}
    \mathcal{L}_T(\iota;\tb,\Lmax) &= \Lmax \begin{cases} 
      1 & \iota \leq \tb \\
      0 & \iota > \tb 
   \end{cases} \hspace{1.9cm} \label{eq:tophat}
\end{align}
\begin{align}
   \mathcal{L}_G(\iota;\tb,\Lmax) &= \Lmax e^{-\frac{\iota^2}{2 \tb^2}} \label{eq:gauss} \\ 
   \mathcal{L}_P(\iota;\tb,\Lmax) &= \Lmax \begin{cases} 
      1 & \iota \leq \frac{\tb}{2} \\
      (\frac{2 \iota}{\tb})^{-2} & \frac{\tb}{2} < \iota \leq \tb \\
      0 & \iota > \tb 
   \end{cases} \label{eq:broken_plaw}
\end{align}
We note that the exponent in Equation \ref{eq:broken_plaw} is fixed, and is chosen for consistency with other studies \citep{wandermanpiran,2002ApJ...571..876Z,kentaro,sylvia}.

\begin{figure*}
    \includegraphics[width=0.5\textwidth, clip=True, trim=0.0cm 0.0cm 0.0cm 0.7cm]{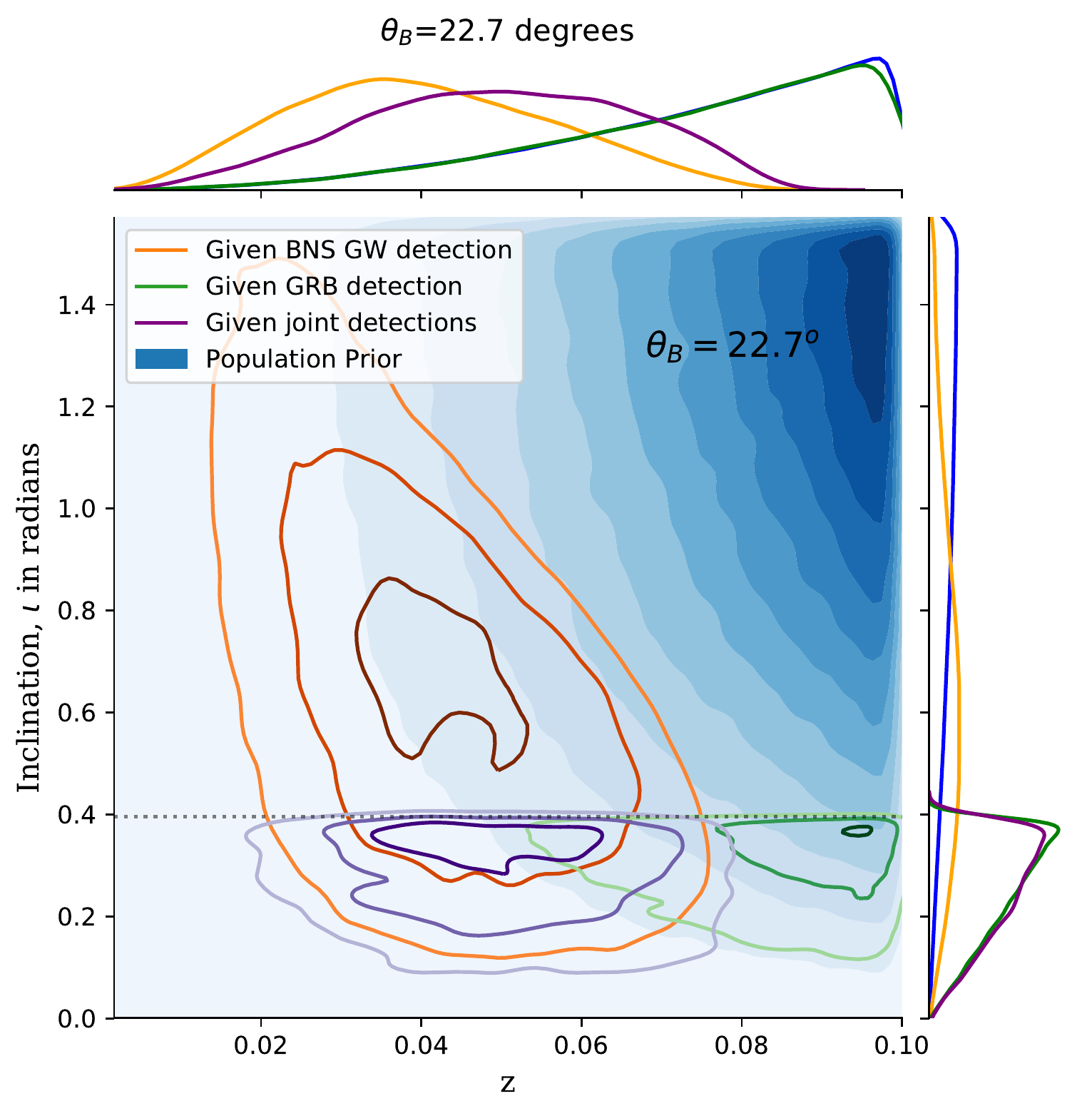}
    \includegraphics[width=0.5\textwidth, clip=True, trim=0.0cm 0.0cm 0.0cm 0.7cm]{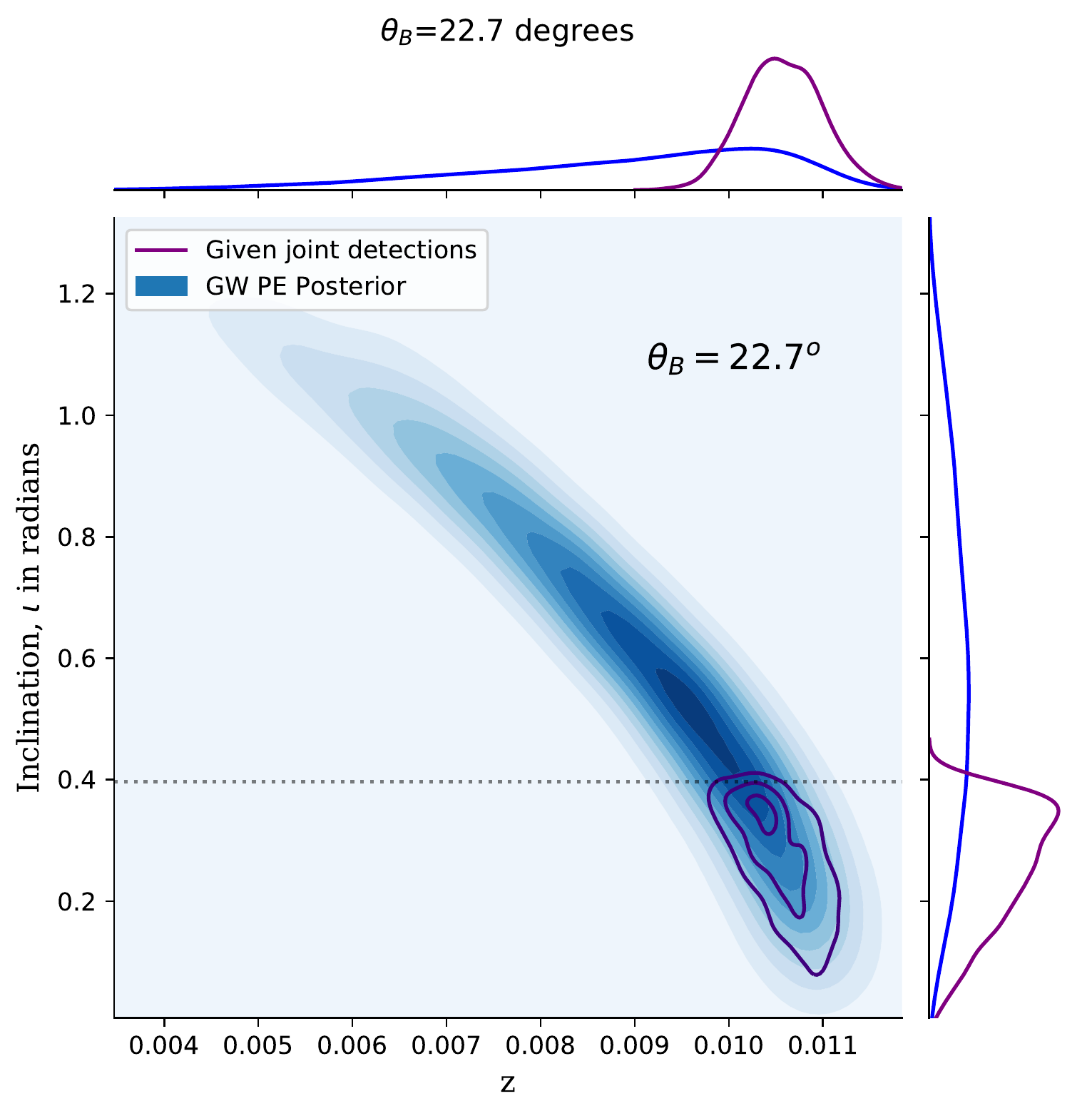}
    \caption{
        Illustration of selection effects with $\mathcal{O}(10^6)$ BNS systems uniformly distributed in comoving volume and isotropically oriented with tophat jets with $\tb=22.7^\circ$.
        (\emph{left}) Distributions without GW parameter estimation, as would be used in simple counting experiments.
        The prior $p(z,\iota)$ is shown in blue and the population of detected GW signals in yellow.
        (\emph{right}) The posterior distribution from GW170817, in which the blue shading acts like the yellow contours in the left panel. 
        Note the different scales on both axes between the two panels.
        (\emph{both})
        The observed luminosity of each event is calculated for various values of $\tb$ (horizontal dashed grey line), and the population detected by GBM is shown with green contours.
        The population of joint detections is shown in purple contours.
        Contour levels are arbitrary and solely for illustrative purposes.
        In comparison to the GW distribution over $z$ and $\iota$ without GW170817 parameter estimation, there is more support at lower inclination and lower redshift with parameter estimation (note that the x-limits in these figures differ by an order of magnitude).
        One would expect higher coincident fractions at a lower beaming angle than for the population prior.
        This can be seen in Figure \ref{fig:q_vs_tb}.
    }
    \label{fig:detected_pop} 
\end{figure*}

Now, we need the distribution of inclinations and redshifts given a GW detection, namely

\begin{align}
    p(\iota,z|\Ab) & = \frac{p(\iota,z,\Ab)}{p(\Ab)} \nonumber \\
                   & = \frac{p(\iota,z)p(\Ab|z,\iota)}{\int \diff z \diff \iota\, p(\Ab|z, \iota) p(\iota,z)} \nonumber
\end{align}
given a GW selection function $p(\Ab|z,\iota)$ and a prior $p(\iota, z)$.
We model the GW selection function assuming a single detector signal to noise ratio threshold of $\rho_{\mathrm{thr}} = 8$.
Thus, $p(\Ab|z,\iota) = \Theta(\rho(z,\iota) - \rho_{thr})$ and we obtain

\begin{equation}
    p(\iota,z|\Ab) =  \frac{\Theta(\rho(z,\iota) - \rho_{thr}) p(\iota,z) }{\int \diff z \diff \iota\, \Theta(\rho(z,\iota) - \rho_{thr}) p(\iota,z)} 
    \label{eq:pdet_bns_heaviside}
\end{equation}
Many factors contribute to an event's signal in a LIGO detector, such as the orientation of the source, distance to the source, and location of the source in the sky. 
A full explanation of these effects can be found in \cite{Finn_1993}.
\textit{A priori}, we expect a BNS's redshift to be completely independent of its inclination with respect to Earth, so $p(\iota,z) = p(\iota)p(z)$.
We further assume that in our detectable redshift range, BNS mergers are distributed uniformly in comoving volume $V_c$, so that 

\begin{align*}
    p(z) &= p(V_c) \frac{\diff V_c}{\diff z} \propto \frac{\diff V_c}{\diff z}
\end{align*} 
We adopt a $\mathrm{\Lambda}$CDM cosmology with $H_0 = 70$ km  sec$^{-1}$ Mpc$^{-1}$ and $\Omega_m=0.3$. It is also safe to assume that the BNS inclinations are isotropically distributed. 
Thus, 

\begin{equation}
    p(\iota,z) \propto \frac{\diff V_c}{\diff z} \sin{\iota}
    \label{eq:p_iz_prior}
\end{equation}
Bringing this all together, we substitute Equations \ref{eq:pdet_grb_heaviside}, \ref{eq:Liso}, and \ref{eq:pdet_bns_heaviside} into Equation \ref{eq:marginalized_q} to obtain
\begin{widetext}
\begin{equation}
    q = \focc \frac{\int \diff z \diff \iota\, p(\iota,z) \Theta(4 \pi \mathcal{L}(\iota;\tb,\Lmax) - L_{\mathrm{iso,thr}}(z)) \Theta(\rho(z,\iota) - \rho_{thr})}{ \int \diff z \diff \iota\, p(\iota,z) \Theta(\rho(z,\iota) - \rho_{thr}) }
    \label{eq:subbed_in_marg_q}
\end{equation}
which we evaluate with Monte-Carlo estimates of each integral, drawing $M = O(10^6)$ samples for $z$ and $\iota$ from Equation~\ref{eq:p_iz_prior} and $\Lmax$ drawn from a Gaussian distribution with mean $2 \times 10^{52}$ erg/s and standard deviation of 10\% of the mean.

\begin{equation*}
    q \approx \frac{\focc \sum\limits_{j=1}^{M} \Theta(4 \pi \mathcal{L}(\iota_j;\tb, \Lmax^{(j)}) - L_{\mathrm{iso,thr}}(z_j))\Theta(\rho(\iota_j,z_j) - \rho_{\mathrm{thr}})}{ \sum\limits_{j=1}^{M} \Theta(\rho(\iota_j,z_j) - \rho_{\mathrm{thr}})} 
\end{equation*}
\end{widetext}
This looks like the ratio of coincident detections to all GW detections.
This is illustrated in Figure \ref{fig:detected_pop}. 
The population of sGRBs that would be detected varies drastically with different $\tb$.
The ratio of joint detections to GW-only detections is determined as a function of the beaming angle.
This ratio, which is the coincident fraction $q(\tb)$, is shown in Figure \ref{fig:q_vs_tb}.

Up until now, we have only considered the number of sGRBs observed in coincidence with a GW event.
We relied on our knowledge of the selection effects of the detectors to provide information on the inclination and redshift of the BNS systems.
However, detected systems provide additional information, since their parameters (such as redshift and inclination) can be estimated from the GW signal.
Our simple counting experiment can be naturally extended to include this additional information, as we now demonstrate. 

\subsection{Including Information from GW Parameter Estimation}
\label{sec:pe}

Consider an individual BNS merger, denoted by $i$, that is detected in GWs with data $d_i$. The coincident fraction for this event ($q_i$) is the likelihood of it being detected in gamma-rays given that it was detected in gravitational waves. We note that this quantity may be different for each event as we condition on different observed GW data. That is,

\begin{align*}
   q_i & = p(\Ag^{(i)} | d_i, \Ab^{(i)} = 1, \tb) \nonumber \\
       & = \frac{p(d_i, \Ag^{(i)}, \Ab^{(i)}=1|\tb)}{p(d_i, \Ab^{(i)}=1)}
 \end{align*}
\begin{widetext}
We note that the numerator can be written as

\begin{align}
    p(d_i, \Ag^{(i)}, \Ab^{(i)}=1|\tb) & = \int \diff z \diff \iota\, p(z, \iota) p(d_i, \Ab^{(i)}=1|z, \iota) p(\Ag^{(i)}|z, \iota, \tb) \nonumber \\
                                       & = \int \diff z \diff\iota\, \left(p(z, \iota|d_i, \Ab^{(i)}=1)p(d_i, \Ab^{(i)}=1)\right) p(\Ag^{(i)}|z, \iota, \tb)
\end{align}
so that

\begin{align}
   q_i =  \int \diff z_i \diff \iota_i\, p(z_i, \iota_i|d_i, \Ab^{(i)}=1) p(\Ag^{(i)} | z_i, \iota_i,\tb)
\end{align}
\end{widetext}

Note that $p(z, \iota|d_i, \Ab^{(i)}=1)$ is simply the GW posterior on inclination and redshift, so we can compute $q_i$ by Monte-Carlo sampling $p(\Ag|z,\iota;\tb)$ from the GW posterior.
We can thus think of $p(z, \iota|d_i, \Ab^{(i)}=1)$ obtained from GW data as our new prior on $z$ and $\iota$ for this event.

Because the overall rate of BNS mergers is unknown, a joint analysis of multiple events would need to explicitly account for the selection effects associated with different possible merger rates throughout the universe.
As is common in the GW literature, we can adopt an inhomgeneous Poisson likelihood
\begin{multline}
    p(\{d_i,\Ab^{(i)},\Ag^{(i)}\}) = \\
    \left[\prod\limits_i^{\Nb} \int dz d\iota\, p(z,\iota) p(d_i|z, \iota, \Ab^{(i)}) p(\Ag^{(i)}|z,\iota) \right] e^{-\mathcal{N}}
\end{multline}
where $\mathcal{N}$ is the expected number of detections given the population described by $p(z,\iota)$ implicitly included in each $q_i$ and an overall merger rate.
By again considering the probability of detecting sGRBs given the knowledge of GW detections, we obtain the following joint likelihood

\begin{align}
    \Lambda & = p(\{\Ag^{(i)}\}|\{d_i,\Ab^{(i)}\}) = \frac{p(\{d_i,\Ab^{(i)},\Ag^{(i)}\})}{p(\{d_i,\Ab^{(i)}\})} \nonumber \\
            & = \frac{\left[\prod \int dz d\iota\, p(z,\iota) p(d_i|z, \iota, \Ab^{(i)}) p(\Ag^{(i)}|z,\iota)\right]e^{-\mathcal{N}}}{\left[\prod \int dz d\iota\, p(z,\iota) p(d_i|z, \iota, \Ab^{(i)}) \right]e^{-\mathcal{N}}} \nonumber \\
            & = \prod\limits_i^{\Nb} q_i
\end{align}
This means that our inference about jet parameters is insensitive to the overall rate of mergers in the universe since we condition on the fact that the system was detected in GWs, although we are still sensitive to the distribution of mergers throughout the universe through the implicit dependence on $p(z,\iota)$ contained in each $q_i$.

As an example, if there were two gravitational wave events and only one of them had a coincident sGRB detection, we would have

\begin{multline}
    \Lambda = p(d_1,\Ag^{(1)} = 1|\Ab^{(1)}=1,\tb) \\
    \times p(d_2,\Ag^{(2)} = 0|\Ab^{(2)}=1,\tb)
\end{multline}
which is a natural generalization of the binomial distribution, as the first factor is analogous to $q$, and the second to $1-q$.


\section{Results}
\label{sec:results}
To date there has only been one BNS event detected in GWs  \citep{0817_ligo}, and it was accompanied by an sGRB~\citep{0817_fermi}.
In this section, we first discuss the implications of this fact using the counting experiment formalism described in Section \ref{sec:q}.
We will then present how these results differ when we take advantage of information provided by both GW parameter estimation and very-long baseline interferometry (VLBI) measurements of inclination \citep{Mooley_2018,superlum_H0}.
Finally, we provide predictions for the future constraining power given various numbers of GW and sGRB detections.

\begin{figure}
    \includegraphics[width=1.0\columnwidth]{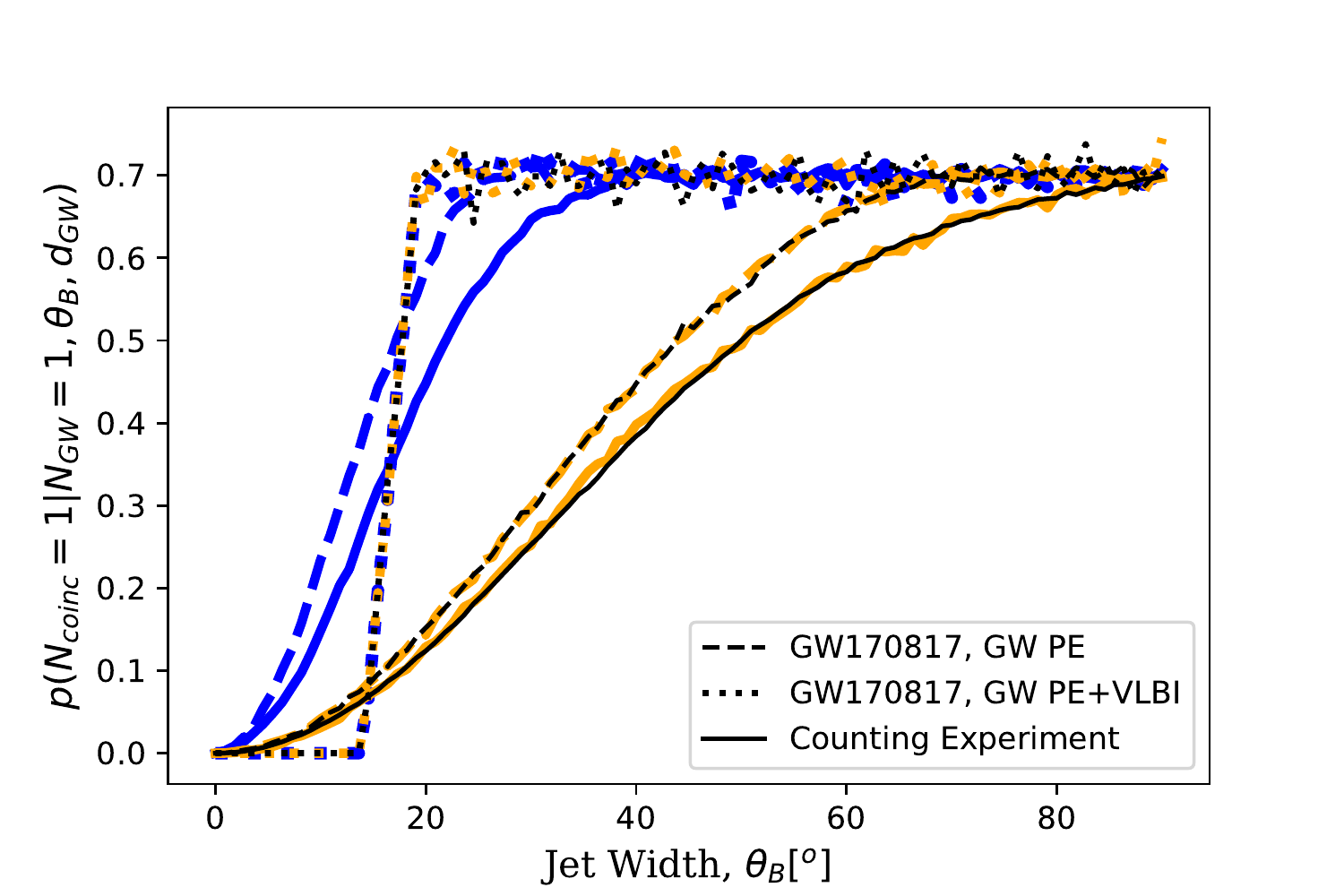}
    \caption{
        Posterior probability distributions for $\tb$ with and without parameter estimation for GW170817.
        Please note that the precise meaning of $\tb$ depends on the jet structure model and constraints on $\tb$ may not be directly comparable between models (see Figure~\ref{fig:dragon}).
        We show the tophat (\emph{black}), broken power law (\emph{orange}), and Gaussian (\emph{blue}) jet structure models with constraints from the counting experiment alone (\emph{solid}), GW parameter estimation (\emph{dashed}), and GW parameter estimation and VLBI measurements (\emph{dotted}).
        We see that the exclusion of small $\tb$ becomes more certain as parameter estimation results provide more precise measurements of the viewing angle.
        Indeed, the measurement using VLBI confidently rules out jet widths less than $14^\circ$.
        The precision of these curves is limited by the finite number of Monte-Carlo samples available, which is particularly evident around the maximum value of the posterior probability for VLBI.
    }
    \label{fig:q_vs_tb}
\end{figure} 


\subsection{Counting Experiment with GW170817}

Given $\Nb = \Ng = 1$, Equation \ref{eq:posterior} simplifies to $p(\tb|\Nb = 1, \Ng = 1) \propto q(\tb)$.
Figure~\ref{fig:q_vs_tb} shows the coincident fraction, proportional to the posterior probability for one event, as a function of $\tb$ based on the counting experiment. 
For all jet structures considered, this rules out $\tb \approx 0$ and favors large jet widths, which is to be expected since $100\%$ of BNS detections have been accompanied by a sGRB. 

For a tophat jet structure, $\tb$ is the angular width of the uniform portion (Equation~\ref{eq:tophat}). 
At inclinations beyond $\tb$, gamma-rays do not reach GBM. 
Since a sGRB was detected, the tophat model has the most support for $\tb = 90^\circ$.
However, because there is significant uncertainty in the viewing angle based on GW selection effects alone, there is rather broad support for many $\tb$.
For the broken power law model, $\tb$ is the inclination at which the luminosity goes to zero (Equation~\ref{eq:broken_plaw}). 
It is also twice the angular width of the uniform core. 
The coincident fraction in this case traces that of the tophat jet, indicating that the truncation at $\tb$ is the dominant feature, rather than the falloff in the region $\tb/2 < \iota < \tb$. 
Recall that the exponent in the power law region was chosen to be $-2$, but larger negative powers up to $-5$ were explored, and did not have noticeable impacts on the results. 
Additionally, truncating the power law at $\iota = 90^\circ$ instead of at $\iota = \tb$ yielded results similar to those of the Gaussian jet.
The differences in constraints on $\tb$ between models, then, is primarily an artifact of how we parameterized the models.
Figure~\ref{fig:dragon} shows the uncertainty in the luminosity distribution itself, and we see that the actual model dependence is much smaller than Figure~\ref{fig:q_vs_tb} might suggest.
For a Gaussian jet structure, recall that we define $\tb$ as the standard deviation (Equation \ref{eq:gauss}). 
Since the Gaussian jet has wide tails, it allows for relatively high luminosities at all inclinations even with a small standard deviation, leading to more support for lower $\tb$. 
If we instead define $\tb$ to be four times the standard deviation, such that the luminosity at $\tb$ is approximately four orders of magnitude lower than at the center of the jet, then the coincident fraction for a Gaussian jet closely traces those of the broken power law and tophat models for equivalent values of $\tb$. 
This is because, for the majority of events seen in GWs, the isotropic luminosity threshold is $\sim 10^{48}$erg/s (Fig.~\ref{fig:sens_curve}), which is approximately four orders of magnitude lower than typical values of $\Lmax$. 
Parameterizing the Gaussian model using $\tb=4\sigma$ instead of $\tb=1\sigma$ does not affect the inference on the jet structure, but would make $\tb$ the point at which the luminosity of the jet crosses GBM's isotropic luminosity threshold, and hence more analogous to the meaning of $\tb$ for the other models.


\subsection{GW170817 GW Posteriors and sGRB Structure}
\label{sec:with pe}

Using parameter estimation from GW170817 obtained from \cite{gwtc} changes this measurement somewhat.
Slightly tighter constraints on inclination angle and redshift nearly rule out that the system is edge-on, so a $90^\circ$ jet opening angle is not as necessary to explain the fact that gamma-rays were observed.
However, $90^\circ$ jets are still consistent with the data.
This allows for slightly more support for medium-width jets ($\tb \sim 45^\circ$--$70^\circ$) than is provided from the simple counting experiment, demonstrated in Figure \ref{fig:q_vs_tb}.
However, this improvement is marginal even for loud events because inclination and distance are degenerate and thus cannot be determined to high precision separately, apart from what is known about their impact on LIGO's sensitivity. 
Only extreme outliers will lead to strong constraints on the inclination from the GW data alone~\citep{Hsin-Yu2019}.
In this way, the counting experiment method is rather powerful because most of the information about a GW event's inclination and redshift is provided by the fact that it was detected. 

When an external measurement is used to break the inclination-distance degeneracy, such as a redshift from identification of a host galaxy, or the inclination from superluminal motion of the radio afterglow of an sGRB \citep{superlum_H0}, constraints are much improved.
In the case of GW170817, \cite{superlum_H0} finds the inclination to be between $\sim14^\circ$ and $\sim19^\circ$.
Given that a sGRB was observed, this implies that the beaming angle must be larger than the lower bound of this region for the tophat case.
The broken power law result traces that of the tophat, and the Gaussian result strictly favors $\tb > 4^\circ$.
Again, these differences reflect parameterization choices.
Figure~\ref{fig:dragon} shows the uncertainty in $\mathcal{L}(\iota)$.

Nonetheless, as with the simple counting experiment, the posteriors on $\tb$ are essentially sigmoids of varying steepness.
Tighter constraints on the inclination for individual events will make these curves somewhat steeper, but their general morphology will remain the same.
We note, then, that combining information from multiple events will be equivalent to multiplying sigmoids.
Detections of sGRBs will exclude small $\tb$ and non-detections will exclude larger $\tb$, producing a narrow window of posterior probability around the true value.
We explore this further in the context of multiple detections below.


\subsection{Projected Constraints with Future Events}

\begin{figure}
    \includegraphics[width=0.5\textwidth, clip=True, trim=0.0cm 1.1cm 0.0cm 0.0cm]{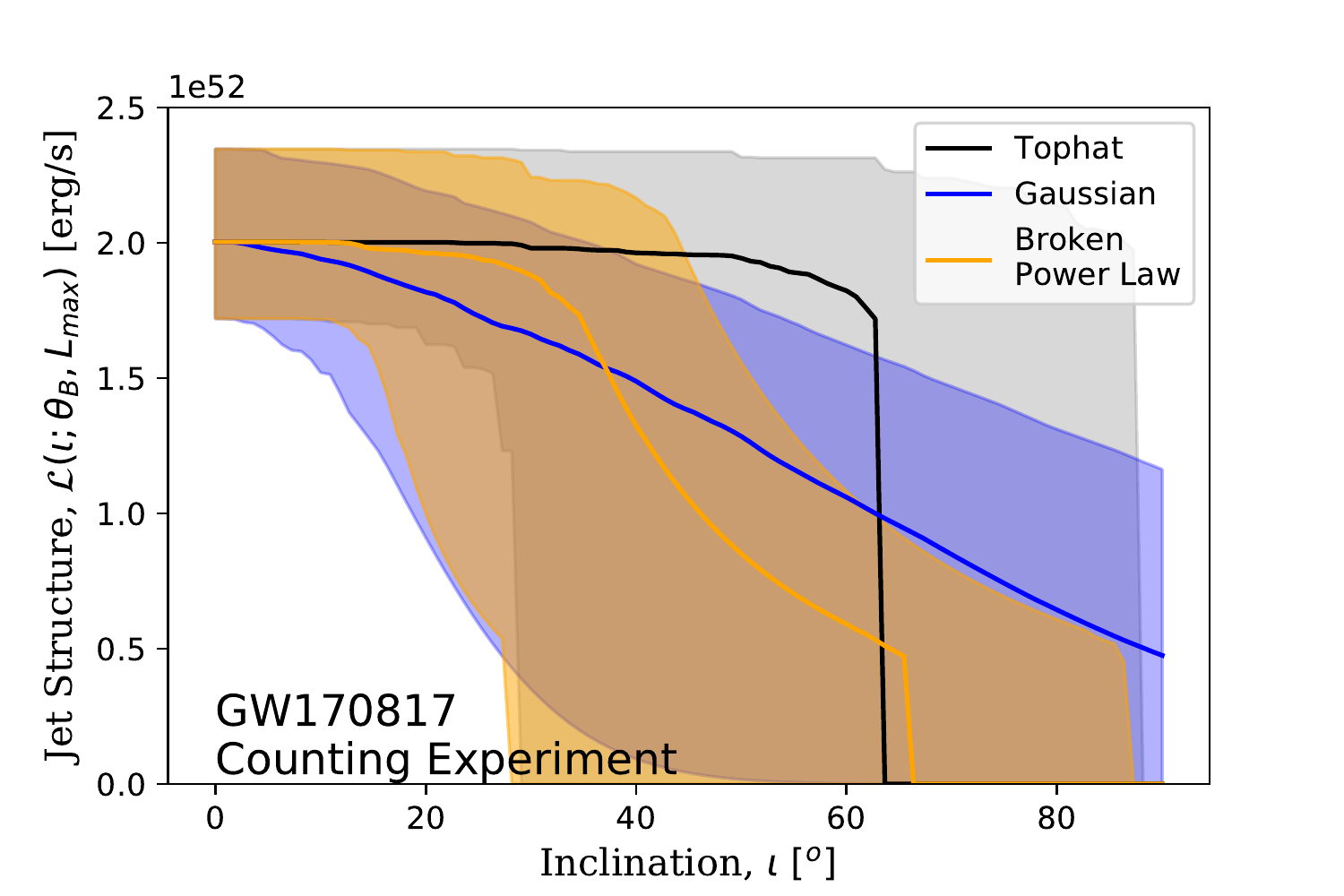} \\
    \includegraphics[width=0.5\textwidth, clip=True, trim=0.0cm 1.1cm 0.0cm 0.0cm]{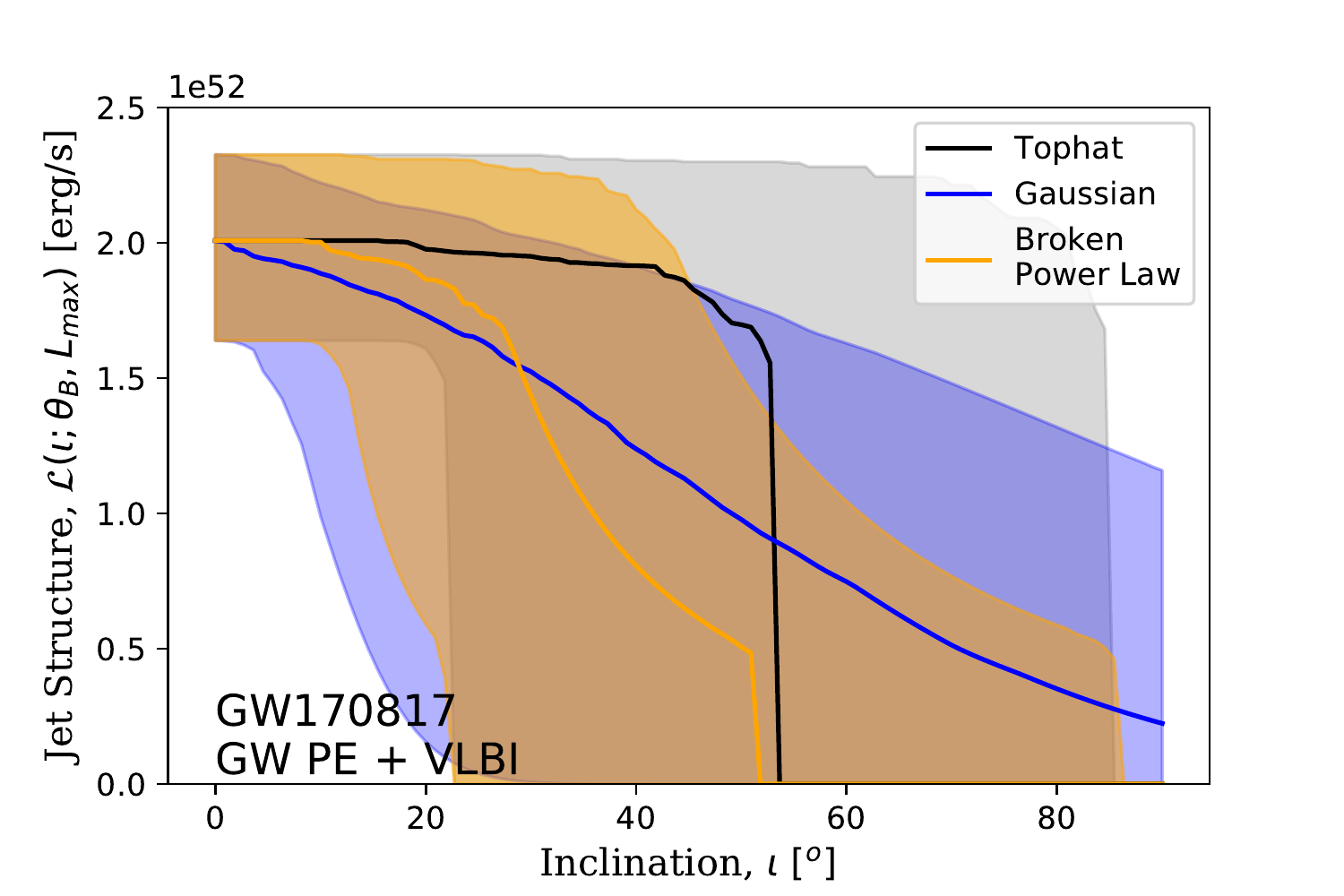} \\
    \includegraphics[width=0.5\textwidth, clip=True, trim=0.0cm 0.0cm 0.0cm 0.0cm]{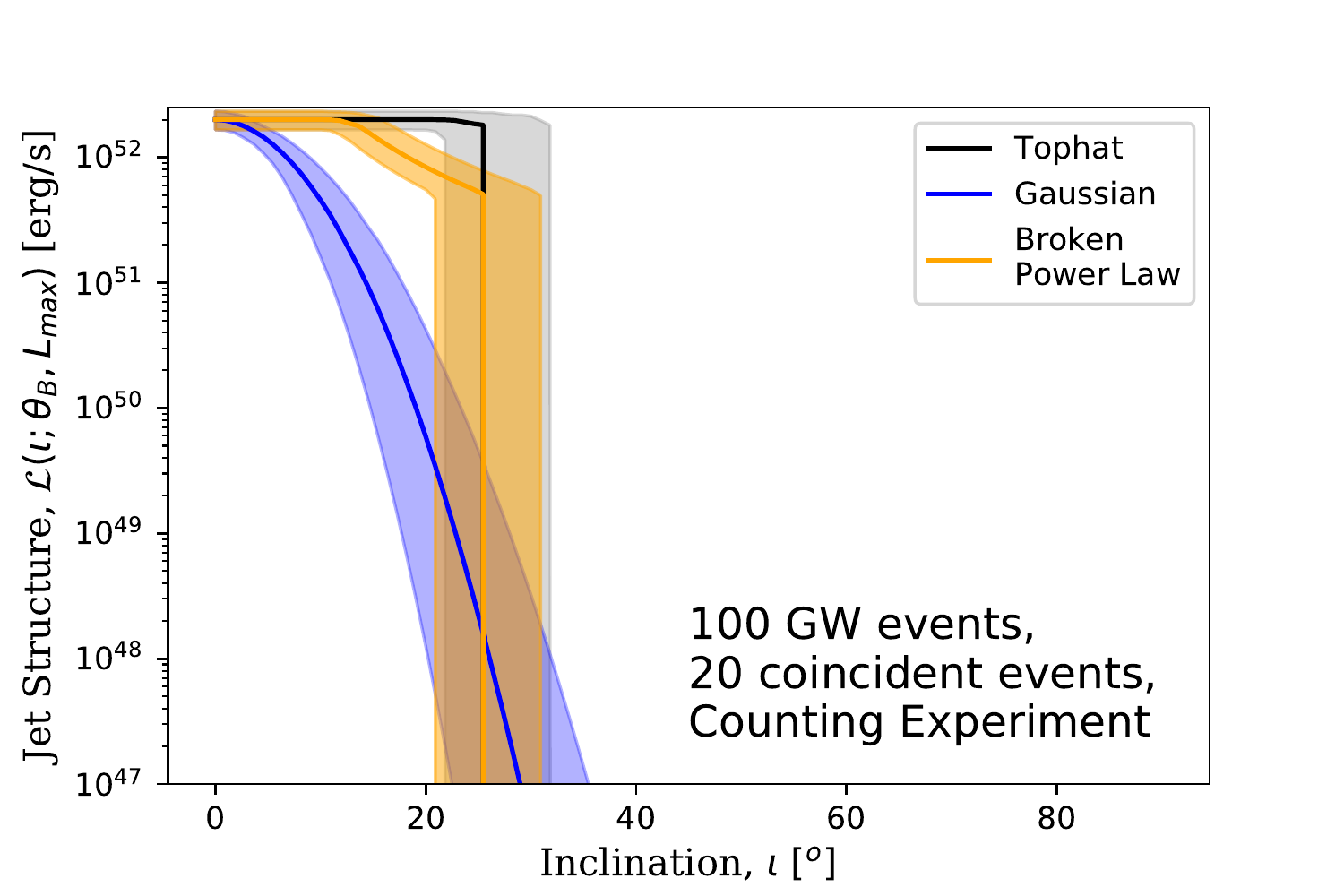}
    \caption{
        Uncertainty in jet structure with the tophat (\emph{black}), broken power law (\emph{orange}), and Gaussian (\emph{blue}) models.
        (\emph{top}) Constraints from the counting experiment with GW170817 alone.
        (\emph{middle}) Constraints from GW170817 including GW parameter estimation and the precise determination of the viewing angle from VLBI measurements.
        (\emph{bottom}) Expected constraints from the counting experiment alone after 100 GW events with 20 coincident sGRBs, shown on a logarithmic rather than a linear y-scale.
        We note that the constraints obtained from GW170817 are nearly identical regardless of whether we perform a simple counting experiment or use the full set of available parameter estimation.
        As such, we only show the expected constraints after 100 GW events with the counting experiment because it is indistinguishable from the expectation with simulated GW parameter estimation.
    }
    \label{fig:dragon} 
\end{figure}

\begin{figure*}
    \includegraphics[width=0.5\textwidth, clip=True, trim=0.0cm 0.0cm 0.0cm 0.7cm]{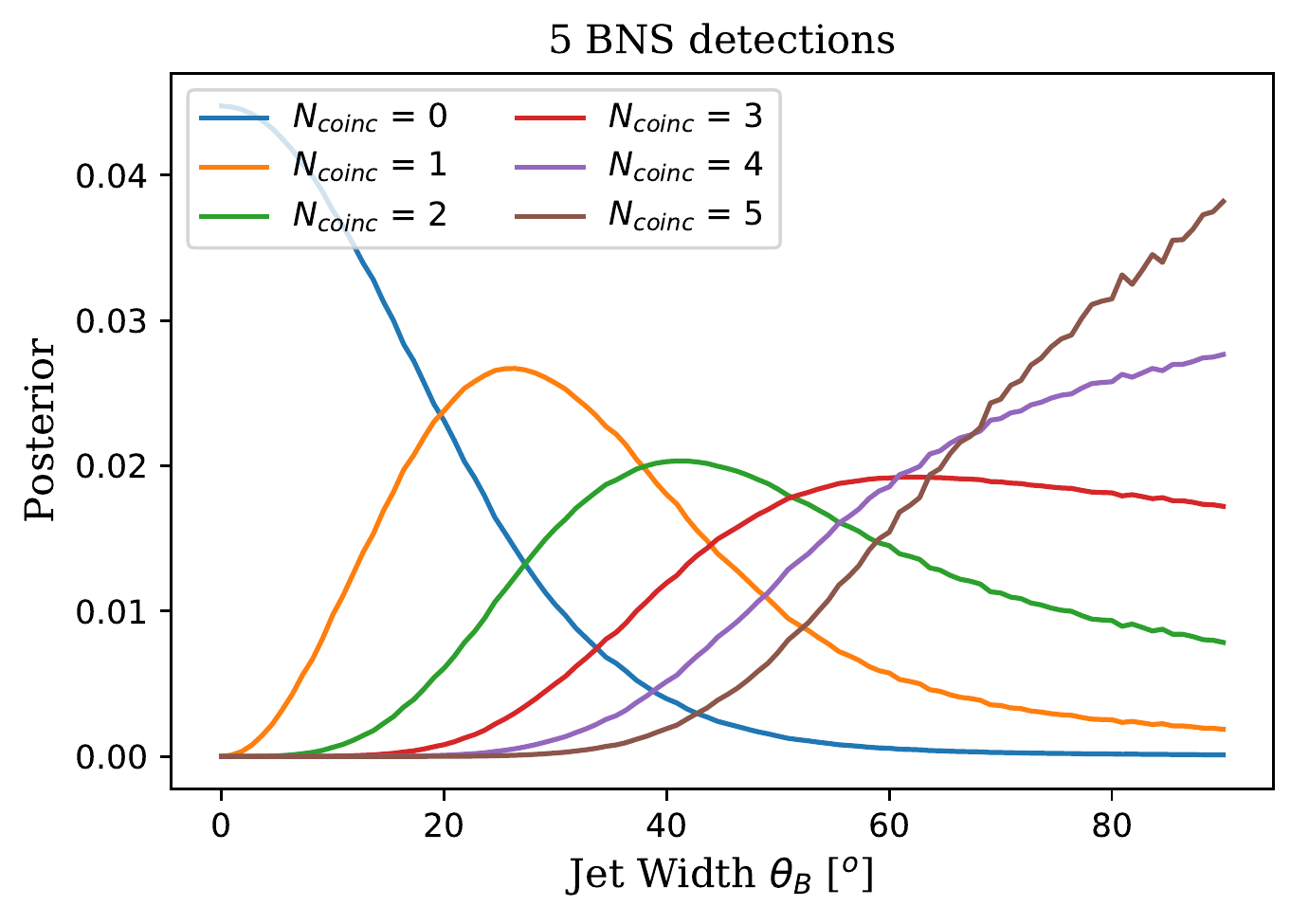}
    \includegraphics[width=0.5\textwidth, clip=True, trim=0.0cm 0.0cm 0.0cm 0.7cm]{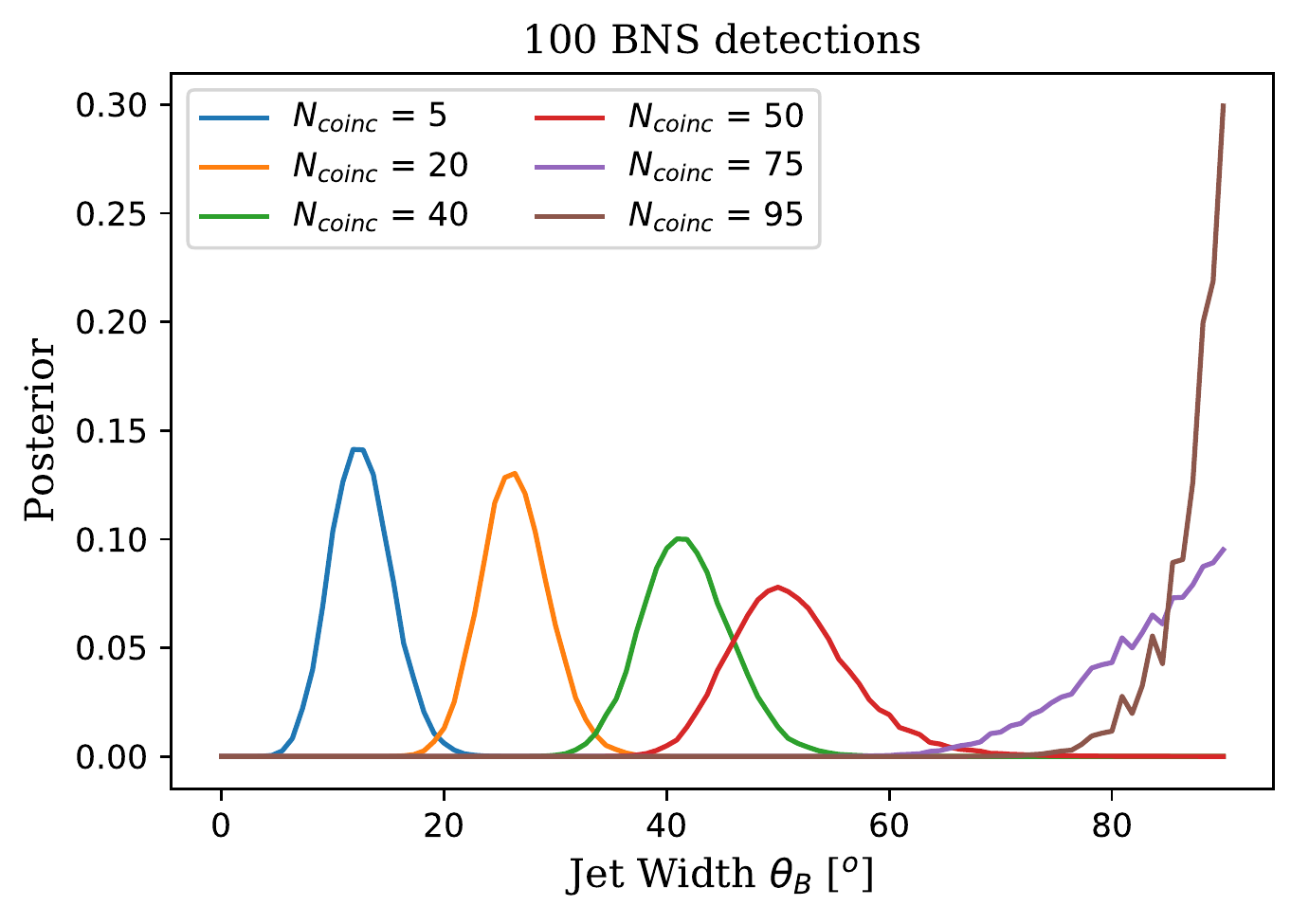}
    \caption{
        Potential future constraints from our counting experiment for (\emph{left}) 5 and (\emph{right}) 100 BNS systems detected in GWs, given the tophat jet model.
        We generally obtain tighter constraints with more events and note that the improvement from GW parameter estimation will become negligible after $\sim100$ BNS detections.
    }
    \label{fig:10+100evs} 
\end{figure*}

In the future, LIGO and Virgo will detect many more BNS systems.
Given the detection of one event in O2 and increased sensitivity in O3, it is possible that several new mergers will be detected by the end of O3~\citep{Aasi:2013wya}.
Thus, we generate posteriors for $\Nb=5$ and various $\Ng$ using the coincident fractions for the three jet structures considered here (Figure \ref{fig:q_vs_tb}).
We note that sources detected in GWs are detected at fairly low redshift, and the mean of the distribution from which $\Lmax$ is drawn is high enough that the $\Lmax$ of any given event is still orders of magnitude higher than the isotropic luminosity threshold for GBM (Figure \ref{fig:sens_curve}).
Thus, jet structure will not bring many events below that threshold, even at high inclination.
As such, we show an example of the constraints on a tophat jet obtained from a counting experiment with 5 and 100 GW detections in Figure~\ref{fig:10+100evs}.
With 100 detections, $\tb$ will be constrained to within $\sigmacounthundred^\circ$ for the tophat model if $\Ng=20$.
We illustrate implications of these constraints on inferred jet structure in Figure~\ref{fig:dragon}. 
In the case of 100 BNS systems detected in GWs, with 20 of them being coincident with sGRBs, we transform the posterior on $\tb$ into a 90\% confidence region of luminosity as a function of inclination for each jet structure model. 

While there is a clear model dependence on these credible intervals within the jet, all these models agree that there is essentially no emission beyond $\sim30^\circ$ in this simulation.
The three models overlap at $\iota \sim 25^\circ$, which corresponds to an isotropic equivalent luminosity of $\sim 10^{48}$ erg/s. 
This is because GBM's isotropic luminosity threshold is close to this value for most sources considered (see Figure \ref{fig:sens_curve}) as most of the simulated events' redshifts are above $z\sim0.01$.
From this, we see that the method outlined in this work primarily constrains the redshift-averaged viewing angle beyond which no sGRBs are observed by GBM, given a detection in gravitational waves.
This angle can be thought of as an ``effective beaming angle," and it depends on the distribution chosen for $\Lmax$, as well as the sensitivity of the instrument used to observe gamma-rays.
Modeling systematics may then affect the inference of jet structure, but they will not strongly influence our first-order conclusions, like whether there is significant emission outside of a narrow window.

\begin{figure}
    \centering
    \includegraphics[width=1.0\columnwidth, clip=True, trim=0.0cm 0.0cm 0.0cm 1.1cm]{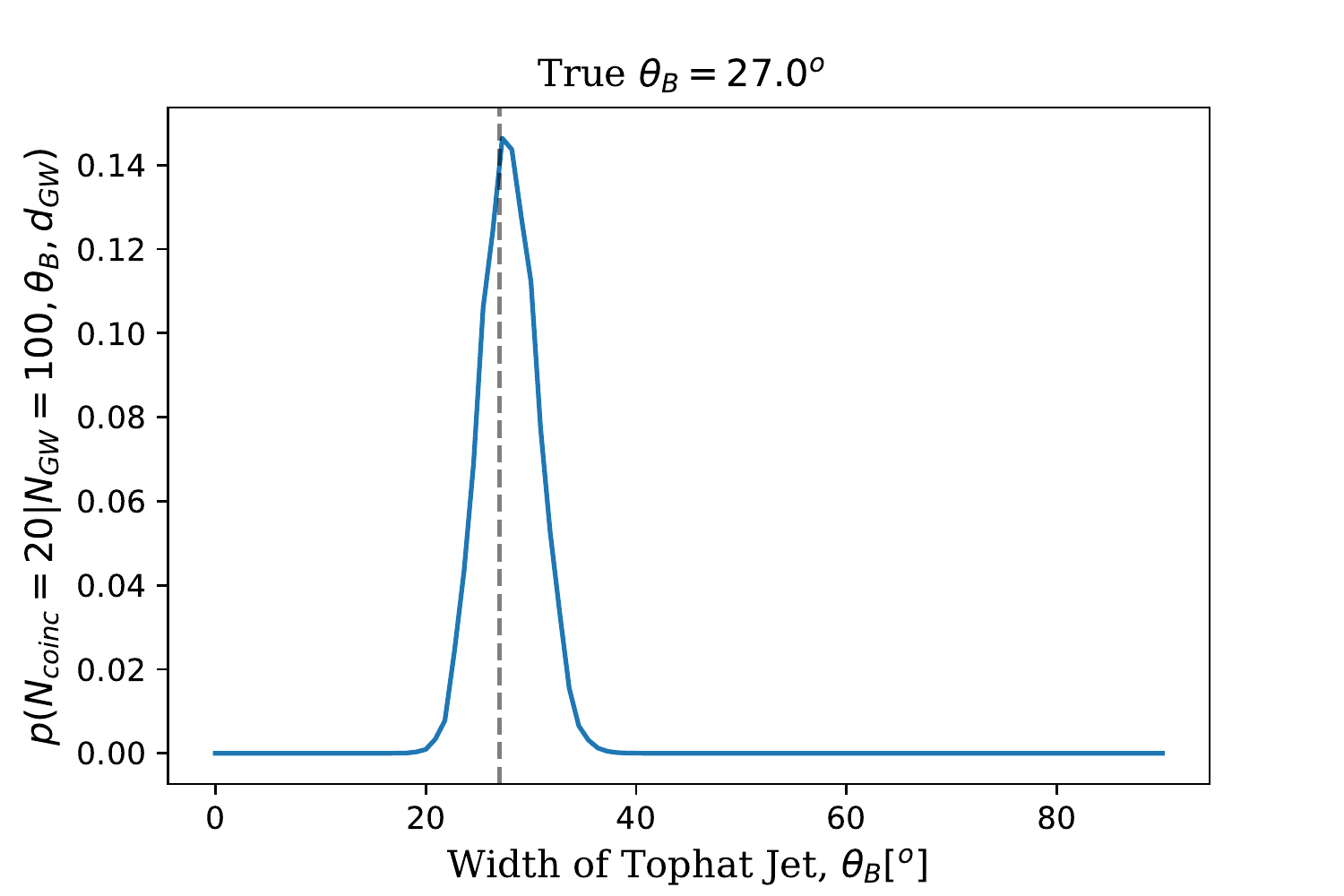}
    \caption{
        Total likelihood with simulated GW parameter estimation for the case of 100 events with a tophat jet.
        In this simulation, $\tb=27^\circ$ (vertical dashed grey line), within the expected uncertainty from the posterior.
        This true value is chosen for easy comparison with the $\Nb=100, \Ng=20$ result from the counting experiment (orange curve in right panel of Figure~\ref{fig:10+100evs}).
        We note that the uncertainty in $\tb$ is comparable to what is achieved by the simple counting experiment with the same number of events.
    }
    \label{fig:mock_pe} 
\end{figure}

Including GW parameter estimation, however, modifies the likelihood.
Although improvements are small in the individual event case, with many events this could significantly impact the overall likelihood of the jet width.
To get a sense of how this will change with real events, mock parameter estimation posteriors are created using the relation between for true inclination and inclination uncertainty presented in Figure 4 of \cite{Hsin-Yu2019}.
We simulate 100 BNS merger events with a range of beaming angles.
This is shown in Figure \ref{fig:mock_pe}.
Note that these projections for future constraints on $\tb$ are conservative in the sense that there are calculated under the assumption that neither VLBI observations nor host galaxy identification are available.

GW parameter estimation can significantly improve the constraints on $\tb$ with only a few events, but it produces nearly identical constraints compared to the simpler counting experiment in the limit of many detections.
This is because the individual event likelihoods are sigmoids, and the product of multiple sigmoids is still a sigmoid.
As discussed in Section~\ref{sec:with pe}, parameter estimation can make the individual event likelihoods steeper than what is produced by the counting experiment, and this leads to an overall tighter constraint, particulary with a small number of events.
However, the joint posterior will be dominated by the single event with the largest (smallest) well-constrained viewing angle that is smaller (larger) than the true $\tb$ and does (does not) have an associated sGRB.
This means constraints will tighten rapidly with the first few events, but will eventually require a large number of detections to obtain an outlier with large (small) enough $\tb$ to further improve our knowledge.

While the counting experiment produces smoother single-event posteriors, we note that the product of many smooth sigmoids is a relatively sharp sigmoid.
Therefore, parameter estimation's initial advantage can be overcome with many events, and the constraints obtained by both approaches will be quite similar after 100 BNS detections.
It is really the overall sensitivities and angular scales associated with GW and sGRB emission and detection, as encoded in simple counting experiments, that drive our ability to constrain the beaming angle with many events.


\section{Discussion}
\label{sec:conclusions}

We present two methods to determine the width of a sGRB jet assuming a specific universal jet structure.
We showed that with 100 BNS detections in GWs, a simple counting experiment will constrain $\tb$ to $\epscounthundred$, and including GW parameter estimation will constrain $\tb$ to within $\epspehundred$ assuming a tophat jet structure.
The methods presented here are complementary to light curve jet break measurements of inclination, and a comparison between the two might provide an interesting test of the existence of a universal jet structure. 
While our counting constraints on $\tb$  may not be extremely informative by the end of O3, compared to those obtained through jetbreak measurements, our method is relatively straightforward, robust, and does not require lengthy electromagnetic observations. It also infers jet structures that are roughly model-independent.
As such, it provides a useful comparison and basic sanity check of our models.

It is also important to emphasize that none of the jet structure models considered here easily produce an event that is observed as sub-luminously as GRB 170817A ($\sim10^{47}$ erg/s) \citep{170817_grb_ligo}.
This is because, as mentioned previously, $\Lmax$ is drawn from a relatively narrow distribution that is peaked at a luminosity orders of magnitude higher than that of GW170817.
While that distribution was picked to be consistent with the majority of observed sGRBs with known redshift, it may still be flawed, and future work will explore the consequences of drawing $\Lmax$ from a wider distribution.
Alternatively, one could simultaneously constrain $\tb$ and $\Lmax$, or choose jet structure models that fall off more steeply with increasing inclination than Gaussian or broken power law models but do not actually vanish.

Another important caveat is that, as mentioned in Section \ref{sec:intro}, in order to do such a calculation one must assume a universal jet structure.
It is likely that this is not the case, as the jet structure could depend on the physical properties of the BNS system and circum-burst environment.
One natural way to account for this would be to constrain the distribution of $\tb$; that is, model each BNS system with a separate $\tb$ but require them all to be drawn from a single distribution.
Such an approach would produce a family of parameterized jet structures and assign a relative likelihood to each:
\begin{equation}
    \Lambda(\lambda) = \prod\limits_{i}^{N_\mathrm{GW}} \int \diff\tb\, p(\tb|\lambda) q_i(\tb),
\end{equation}
where $p(\tb|\lambda)$ describes the relative probability of obtaining different $\tb$, described by the parameters $\lambda$.
However, this also requires us to assume a particular parametrization of the jet structure, which may be difficult to extract from simulations.
If the fidelity of parameterized models is difficult to verify, one might pursue a nonparametric approach instead.
In such approaches, no specific functional form is assumed for the jet structure but instead we rely on the data alone to determine correlations between the luminosity at various viewing angles.
Gaussian processes provide a natural formalism for such an inference scheme and have been pursued in similar contexts within the GW literature~\citep[e.g.][]{reedandphil,2019arXiv191009740E}.
Specifically, one would replace the implicit distribution over $\mathcal{L}$ described by $p(\tb|\lambda)$ with a Gaussian process for $\mathcal{L}$.
If there is a universal jet structure, the nonparametric posterior process for $\mathcal{L}$ should collapse to a single curve, regardless of the true jet structure's functional form.
If the jet structure is not universal, then the nonparametric posterior process will capture the full variability of the jets produced in nature, again regardless of their precise functional form.
Similar techniques have been proposed for parameterized tests of general relativity with GW events~\citep{2019PhRvL.123l1101I}.
Such a nonparametric analysis would avoid modeling errors associated with choosing a functional form for the jet structure \textit{a priori}, and instead would {\em infer the full distribution of jet structures observed in nature directly from the data}.
However, we note that sampling from such nonparametric posterior processes can be nontrivial and leave further exploration to future work.

At present, statistical uncertainty dominates systematic modeling errors with GW170817.
We find that the majority of the constraining power of GWs for sGRB jet widths results from the relative sensitivity to sources at different redshift and inclination in both GW detectors and GBM.
We demonstrate that our simple counting experiment can determine the jet width to nearly the same precision as more complex approaches that rely on computationally expensive parameter estimation techniques. 
We also illustrate the simple extension of our approach to use gravitational wave and electromagnetic parameter estimation, which will more quickly constrain the geometry of sGRBs with a small number of detections but will not significantly improve over counting experiments after $\sim100$ BNS detections.
As such, we can count on sGRBs detected in coincidence with GW events to help us constrain the angular emission profile of all sGRBs, from which we can develop a better understanding of the physical processes driving sGRB jets and the remnants of compact binary mergers.


\section*{Acknowledgements}
The authors are grateful to 
Edo Berger, Cecelia Chirenti, Ryan Foley, Wen-Fai Fong, Chris Fryer, Dan Kasen, Raffaella Margutti, Brian Metzger,  Cole Miller, and Karelle Siellez for helpful discussions about gamma-ray bursts. 
We also thank Karelle Siellez and Eric Thrane for useful conversations about GBM, and for sharing methodologies used to compute GBM's isotropic luminosity threshold.
We thank the LIGO Scientific Collaboration and Virgo Collaboration for public access to data products. 
This research was conducted in part at the Kavli Institute for Theoretical Physics, supported in part by the National Science Foundation under Grant No. NSF PHY-1748958.
AF, RE, MF, ZD, and DEH were supported by the Kavli Institute for Cosmological Physics at the University of Chicago through NSF grant PHY-1125897 and an endowment from the Kavli Foundation.
AF, MF, ZD, and DEH were also supported by NSF grant PHY-1708081.
MF and ZD were supported by the NSF Graduate Research Fellowship Program under grants DGE-1746045 and DGE-1144082, respectively.
DEH also gratefully acknowledges support from the Marion and Stuart Rice Award.


\bibliographystyle{aasjournal}
\bibliography{bibfile}{}

\end{document}